\documentclass[journal]{IEEEtran}
\usepackage{epsfig,graphicx,subfigure,psfrag,amsmath,cases}
\usepackage{latexsym,amsmath,amssymb,epsfig,subfigure,algorithm,mathtools}
\usepackage{algorithmic}
\usepackage{color}
\usepackage{url}
\usepackage{scrtime}
\usepackage{cite}
\usepackage{subfigure}
\usepackage{bbding}
\usepackage{multicol}

\author{{Zhiqiang Wei,~\IEEEmembership{Member,~IEEE,} Shuangyang Li,~\IEEEmembership{Member,~IEEE,} Weijie Yuan,~\IEEEmembership{Member,~IEEE,} Robert Schober,~\IEEEmembership{Fellow,~IEEE,} and Giuseppe Caire,~\IEEEmembership{Fellow,~IEEE}}\\
{(\emph{Invited Paper})}\vspace{-10mm}
\thanks{Z. Wei is with the School of Mathematics and Statistics, Xi'an Jiaotong University, Xi'an 710049, China (e-mail: zhiqiang.wei@xjtu.edu.au).}
\thanks{S.~Li was with the School of Electrical Engineering and Telecommunications, University of New South Wales, Sydney, NSW
	2052, Australia, when this letter was submitted. He is now with the Department of Electrical, Electronic, and Computer Engineering, University of Western Australia, Perth, WA 6009, Australia (e-mail: shuangyang.li@uwa.edu.au).}
\thanks{
	W.~Yuan is with the Department of Electrical and Electronic Engineering,
	Southern University of Science and Technology, Shenzhen 518055, China
	(e-mail: yuanwj@sustech.edu.cn).}
\thanks{R.~Schober is with the Institute for Digital Communications (IDC), the Friedrich-Alexander University Erlangen-Nuremberg, Erlangen 91054, Germany (e-mail: {robert.schober}@fau.de).}
\thanks{G. Caire is with the Electrical Engineering and Computer Science Department, the Technische Universit{\"a}t Berlin, Berlin 10587, Germany (e-mail: caire@tu-berlin.de).}
}

\title{Orthogonal Time Frequency Space Modulation -- Part I: Fundamentals and Challenges Ahead}

\newtheorem{Def}{Definition}

\newtheorem{T-Prob}{Transformed Problem}

\begin{document}
\maketitle
\begin{abstract}
This letter is the first part of a three-part tutorial on orthogonal time frequency space (OTFS) modulation, which is a promising candidate waveform for future wireless networks.
This letter introduces and compares two popular implementations of OTFS modulation, namely the symplectic finite Fourier transform (SFFT)- and discrete Zak transform (DZT)-based architectures.
Based on these transceiver architectures, fundamental concepts of OTFS modulation, including the delay-Doppler (DD) domain, DD domain information multiplexing, and its potential benefits, are discussed. 
Finally, the challenges ahead for OTFS modulation are highlighted.
Parts II and III of this tutorial on OTFS modulation focus on transceiver designs and integrated sensing and communication (ISAC), respectively.
\end{abstract}


\vspace{-2mm}
\section{Introduction}
Next-generation wireless networks are required to support new applications and challenging deployment scenarios, including mobile communications on board aircraft (MCA), low-earth-orbit (LEO) satellites\cite{CHENG2021}, high speed trains, unmanned aerial vehicles (UAVs)\cite{CaiUAVEESecure}, and vehicle-to-vehicle (V2V) networks\cite{WeijieISAC}.
In this context, high-mobility communications have become an essential technology for future wireless networks\cite{wei2020orthogonal}.
Wireless communications in high-mobility propagation environments experience fast time-varying channels, which are typically caused by the Doppler effect\cite{hlawatsch2011wireless}.
{If the Doppler effect was not considered for system design, wireless communications over high-mobility channels would suffer from significant performance degradation.
The widely-adopted orthogonal frequency division multiplexing (OFDM) modulation may not be able to support efficient and reliable communications in high-mobility environments due to the severe inter-carrier interference (ICI) induced by Doppler spread.
Besides, the limited coherence time of the time-frequency domain channel will incur a large channel estimation overhead for OFDM systems.
For example, for an OFDM system operating at a carrier frequency of $f_c = 3.5$ GHz, employing a subcarrier spacing of $\Delta f = 15$ kHz, and supporting a relative velocity of $v = 300$ km/h, the maximum Doppler shift is $\nu_{\mathrm{max}} = 972.22$ Hz while the OFDM symbol duration including a 20\% cyclic prefix (CP) is 80 ms. The channel's
coherence time is around $\frac{1}{4\nu_{\mathrm{max}}} = 257.14$ ms.
Thus, one channel coherence
interval can only accommodate at most 3 OFDM symbols, i.e., a pilot symbol needs to be inserted once every three OFDM symbols, causing a large overhead.}

Recently, a new waveform, namely orthogonal time frequency space (OTFS) modulation, was proposed to cater to high-mobility applications and has attracted significant attention since its introduction\cite{Hadani2017orthogonal,wei2020orthogonal}.
In particular, for OTFS modulation, the information symbols are placed in the delay-Doppler (DD) domain, wherein the time-varying channels are quasi-static and sparse\cite{Hadani2018otfs}.
Besides, DD domain information multiplexing simplifies the coupling between the information symbols and the channels, which facilitates efficient and reliable transceiver design.
More importantly, owing to the two-dimensional (2D) spreading inherent to OTFS modulation, each DD domain symbol experiences the entire time-frequency (TF) domain channel response across one OTFS frame, which enables OTFS to exploit the full degrees of freedom (DoFs) of high-mobility channels.
Moreover, delay and Doppler are typical sensing parameters and the DD domain channel directly mirrors the geometry of the scatterers in high-mobility environments\cite{WeijieISAC,GaudioISAC}.
Therefore, OTFS modulation is regarded as a promising waveform for integrated sensing and communication (ISAC) systems\cite{ShuangyangISAC}.

\begin{table*}[t]
	
	\scriptsize
	{{\caption{Notations for Main System Parameters.}\label{notations}
			\vspace{-5mm}
			\begin{center}
				\begin{tabular}{ c | c |c |c }
					\hline			
					Notation   & Physical meaning                & Notation   & Physical meaning \\ \hline
					${h}_{\mathrm{DD}}\left(\tau,\nu\right)$ & Continuous DD domain channel response & ${H}_{\mathrm{DD}}\left[l,k,l',k'\right]$ & Discrete DD domain channel response\\
					$w\left({l,k, l',k',l_{i},k_{i}}\right)$ & Sampling function & $\alpha\left[l,k,l_i,k_i\right]$ & Phase rotation term \\
					$\mathcal{DZ}_x\left[l,k\right]$  & Discrete Zak transform of $x$  & ${{{A_{{g_{\rm tx}}{g_{\rm rx}}}}\left(\tau,\nu \right)}}$  & Ambiguity function \\\hline
				\end{tabular}
	\end{center}}}
\vspace{-8mm}
\end{table*}

The research on OTFS modulation, including the derivation of its input-output characteristics\cite{RavitejaOTFS,FarhangCPOFDM,MohammedZTOTFS,lampel2021orthogonal},  performance analysis\cite{SurabhiDiversityAnalysisBER,li2020performance,RezazadehReyhaniCPOFDM,ChongOTFS}, transceiver design\cite{RavitejaOTFSCE,WeiWindowOTFS,ZhiqiangOffGrid,ShuangyangCrossDomain2021,ZhengdaoDetection2022}, and its interplay with other promising techniques\cite{ZhiguoNOMAOTFS,gaudio2020hybrid}, is quickly evolving and attracts an increasing number of researchers.
However, it might be challenging for new researchers entering this field to grasp the relevant mathematical tools and abstract concepts.
The authors of \cite{wei2020orthogonal} have provided an overview of the basic concepts, transceiver architectures, and waveforms of OTFS modulation, but did not provide any mathematical details.
This motivates this three-part tutorial, which aims to provide an accessible introduction for new researchers interested in OTFS modulation.
In particular, Part I introduces two popular implementations of OTFS modulation, reviews some related fundamental concepts, and discusses future research challenges.
Part II provides an in-depth discussion of OTFS transceiver design, which is a key research direction in the OTFS literature.
Part III is dedicated to OTFS-based ISAC systems, which are regarded as an enabling technology for the sixth-generation (6G) wireless networks, and highlights a range of general research directions for OTFS modulation.

We note that there are alternative waveforms \cite{TharajOTSM,ZemenOrthogonalPrecoding}, which are closely related to OTFS modulation and can also facilitate high-mobility communications.
However, in this tutorial, we focus on OTFS modulation due to the rapidly growing interest in this modulation format.

\textit{Notations:} 
	$\mathbb{C}^{M\times N}$, $\mathbb{R}^{M\times N}$, and $\mathbb{Z}^{M\times N}$ denote the sets of $M\times N$ matrices with complex, real, and integer entries, respectively;
	$\left(\cdot\right)^*$ denotes the conjugate operation; $\left(\cdot\right)_N$ denotes the modulus operation with respect to (w.r.t.) $N$;	
	$\delta\left(\cdot\right)$ and $\delta\left[\cdot\right]$ are the continuous- and discrete-time delta functions, respectively.
{For clarity, we summarize the notations for the main system parameters adopted in this paper in Table \ref{notations}.}

\vspace{-2mm}
\section{Two Implementations of OTFS Modulation}
{The two most commonly used implementations of OTFS modulation are symplectic finite Fourier transform (SFFT)- \cite{RavitejaOTFS,FarhangCPOFDM} and discrete Zak transform (DZT)-based OTFS \cite{MohammedZTOTFS,lampel2021orthogonal}.}
In this section, we review the details of both transceiver architectures and discuss their similarities and differences.

\vspace{-4mm}
\subsection{SFFT-based OTFS}
\begin{figure}[t]
	\centering
	\includegraphics[width=3.3in]{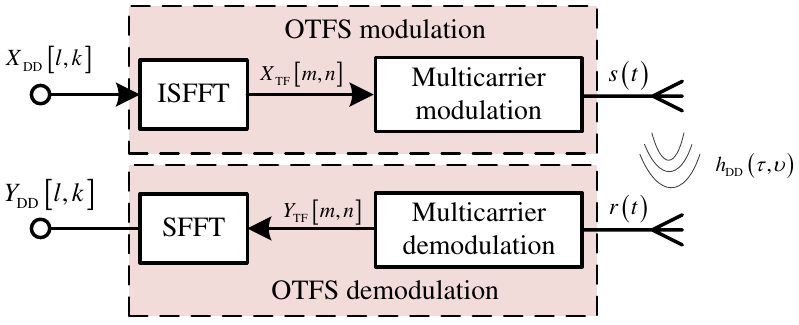}\vspace{-3mm}
	\caption{SFFT-based OTFS transceiver architecture.}
	\vspace{-6mm}
	\label{SFFT_Based_OTFS}%
\end{figure}
The SFFT-based OTFS transceiver architecture is shown in Fig. \ref{SFFT_Based_OTFS}\cite{RavitejaOTFS,FarhangCPOFDM}.
We assume a bandwidth of $B_{\mathrm{OTFS}}$ and a time duration of $T_{\mathrm{OTFS}}$, defined in the TF domain, are available for accommodating one OTFS frame.
Bandwidth $B_{\mathrm{OTFS}}$ is divided into $M$ subcarriers with subcarrier spacing $\Delta f = \frac{B_{\mathrm{OTFS}}}{M}$ and time duration $T_{\mathrm{OTFS}}$ is divided into $N$ time slots with slot duration $T = \frac{T_{\mathrm{OTFS}}}{N}$.
At the transmitter (TX) side, the information symbol for the $\left(l,k\right)$-th DD resource element (DDRE) is denoted as $X_{\mathrm{DD}}\left[ {l,k} \right]$, where $l \in \{0,\ldots,M-1\}$ is the delay index and $k \in \{0,\ldots,N-1\}$ is the Doppler index.
The DD domain symbols $X_{\mathrm{DD}}\left[ {l,k} \right]$ are transformed into TF domain signal $X_{\mathrm{TF}}\left[ {m,n} \right]$ via the inverse symplectic finite Fourier transform (ISFFT) \cite{WeiWindowOTFS}, i.e., 
\vspace{-2mm}
\begin{equation} \label{ISFFT}
X_{\mathrm{TF}}\left[ {m,n} \right] = \frac{1}{{\sqrt {NM} }}\sum\limits_{k = 0}^{N - 1} {\sum\limits_{l = 0}^{M - 1} {X_{\mathrm{DD}}\left[ {l,k} \right]{e^{j2\pi \left( {\frac{{nk}}{N} - \frac{{ml}}{M}} \right)}}} },\vspace{-2mm}
\end{equation}
where $m \in \{0,\ldots,M-1\}$ is the subcarrier index and $n \in \{0,\ldots,N-1\}$ is the slot index.
The OTFS transmit signal ${s}\left(t\right)$ is obtained by performing multicarrier modulation on the TF domain signal ${X}_{\mathrm{TF}}\left[ {m,n} \right]$, also known as Heisenberg transform \cite{RavitejaOTFS}, i.e., 
\vspace{-2mm}
\begin{equation} \label{MulticarrierMod}
s\left( t \right) = \sum\limits_{n = 0}^{N - 1} {\sum\limits_{m = 0}^{M - 1} {X_{\mathrm{TF}}\left[ {m,n} \right]{{g_{{\rm{tx}}}}\left( {t - nT} \right){e^{j2\pi m\Delta f\left( {t - nT} \right)}}} }},\vspace{-2mm}
\end{equation}
where ${{g_{{\rm{tx}}}}\left( {t} \right)}$ is the TX pulse shaping filter.
{In practice, the TF domain signal $X_{\mathrm{TF}}\left[ {m,n} \right]$ is first transformed to a time domain sequence via the inverse finite Fourier transform (IFFT), is then pulse shaped in the discrete time domain, and is finally transformed to the analog TX signal $s\left( t \right)$ by digital-to-analog conversion (DAC).}

A deterministic channel model in the DD domain has been widely adopted for linear time-varying (LTV) channels \cite{hlawatsch2011wireless,hong2022delay} and the continuous DD domain channel response is modeled as
\vspace{-2mm}
\begin{equation}\label{DDDomainChannel}
{h}_{\mathrm{DD}}\left(\tau,\nu\right) = \sum\nolimits_{i=1}^{P} {h_i}   \delta\left(\tau - \tau_i\right)\delta\left(\nu - \nu_i\right),\vspace{-2mm}
\end{equation}
where $P \in \mathbb{Z}$ is the number of resolvable paths and ${h}_i \in \mathbb{C}$ is the channel coefficient of the $i$-th path.
Variables ${\tau_i} \in \left[0, \tau_{\mathrm{max}}\right]$ and ${\nu_i} \in \left[-\nu_{\mathrm{max}}, \nu_{\mathrm{max}}\right]$ denote the delay and Doppler shifts of the $i$-th path, respectively, where $\tau_{\mathrm{max}}$ and $\nu_{\mathrm{max}}$ denote the maximum delay and Doppler shifts in the considered high-mobility propagation environment, respectively.
{{Note that DD domain channel responses fluctuate much slower than time-delay domain or TF domain channel responses, since only a drastic change of the propagation path lengths or the moving speeds would cause DD domain channel response variations.}}

After passing through the LTV channel, the received signal at the OTFS receiver (RX) is given by
\vspace{-2mm}
\begin{align}\label{LTVChannel}
r\left( t \right) &= \int_{-\infty}^{\infty} {\int_{-\infty}^{\infty} {h_{\mathrm{DD}}\left( {\tau,\nu } \right)} } {{e^{j2\pi \nu \left( {t - \tau } \right)}}}s\left( {t - \tau } \right)\mathrm{d}\tau \mathrm{d}\nu + z\left(t\right)\notag\\[-1mm]
& = \sum\nolimits_{i=1}^{P}{h_i}s\left( {t - \tau_i } \right){{e^{j2\pi \nu_i \left( {t - \tau_i } \right)}}} + z\left(t\right),
\end{align}
\vspace{-7mm}
\par
\noindent
where $z\left(t\right)$ denotes the additive white Gaussian noise (AWGN) in the time domain.
At the OTFS RX, after receive filtering, we first perform multicarrier demodulation of $r\left( t \right)$ to obtain the TF domain received signal.
This operation is also known as Wigner transform\cite{RavitejaOTFS} and given by
\vspace{-2mm}
\begin{equation}\label{MulticarrierDeMod}
{Y}_{\mathrm{TF}}\left[m,n\right] = \int_{-\infty}^{\infty} {r\left( t \right)g_{{\rm{rx}}}^ * \left( {t - nT} \right){e^{ - j2\pi m\Delta f\left( {t - nT} \right)}}\mathrm{d}t},\vspace{-3mm}
\end{equation}
where ${{g_{{\rm{rx}}}}\left( {t} \right)}$ is the RX filter.
Then, the TF domain received signal is transformed into the DD domain by the SFFT. The resulting DD domain received signal is given by
\vspace{-3mm}
\begin{equation} \label{SFFT}
Y_{\mathrm{DD}}\left[ {l,k} \right] \hspace{-1mm}=\hspace{-1mm} \frac{1}{{\sqrt {NM} }}\sum\limits_{n = 0}^{N - 1} {\sum\limits_{m = 0}^{M - 1} {{Y}_{\mathrm{TF}}\left[ {m,n} \right]{e^{-j2\pi \left( {\frac{{kn}}{N} - \frac{{lm}}{M}} \right)}}} }.\vspace{-3mm}
\end{equation}
{Note that \eqref{MulticarrierMod} and \eqref{MulticarrierDeMod} degenerate to conventional OFDM modulation and demodulation, respectively, when $N = 1$ and rectangular TX and RX filters are adopted. Moreover, when $N = 1$ and no restrictions on the form of the TX and RX filters are imposed, \eqref{MulticarrierMod} and \eqref{MulticarrierDeMod} degenerate to pulse-shaped OFDM modulation and demodulation, respectively \cite{StrohmerPSOFDM}.}

\setcounter{equation}{8}
\begin{figure*}
	\begin{align}\label{WindowDDDomainRectPulse}
	w\left({l,k, l',k',l_{i},k_{i}}\right)
	&= \frac{1}{{NM}}\sum\nolimits_{n = 0}^{N - 1} \sum\nolimits_{m = 0}^{M - 1} \left[\sum\nolimits_{m' \ne m}^{M - 1} {A_{{g_{\rm tx}}{g_{\rm rx}}}}\left( { - {l_{i} }\frac{1}{{M\Delta f}},\left( {m - m'} \right)\Delta f - {k_{i} }\frac{1}{{NT}}} \right) \right.\notag\\
	&\hspace{-33mm}\left.+ {{e^{ - j2\pi \frac{{k'}}{N}}}\sum\nolimits_{m' = 0}^{M - 1} {{A_{{g_{\rm tx}}{g_{\rm rx}}}}\left( {T - {l_{i} }\frac{1}{{M\Delta f}},\left( {m - m'} \right)\Delta f - {k_{i} }\frac{1}{{NT}}} \right)} }\right] {e^{ - j2\pi \frac{{n\left( {k - k' - {k_{i} }} \right)}}{N}}}{e^{ j2\pi \frac{{\left( {ml - m'l' - m'{l_{i} }} \right)}}{M}}}
	\end{align}
	\vspace{-7mm}
\end{figure*}
\setcounter{equation}{11}
\begin{figure*}
	\begin{equation}\label{PhaseRotation}
	\alpha\left[l,k,l_i,k_i\right]	= \left\{ {\begin{array}{*{20}{c}}
		{e^{j2\pi \frac{\left(l-l_{i}\right)_M k_{i}}{MN}}}&{l\in \left\{l_{i},\ldots,M-1\right\}}\\
		{e^{j2\pi \frac{\left(l-l_{i}\right)_M k_{i}-k_{i}M-\left(k - k_{i}\right)_NM}{MN}}}&{l\in \left\{0,\ldots,l_{i}-1\right\}}
		\end{array}} \right..\vspace{-2mm}
	\end{equation}
	\vspace{-6mm}
	\hrulefill
\end{figure*}
\setcounter{equation}{6}

Combining \eqref{ISFFT}-\eqref{SFFT}, the DD domain input-output relationship for OTFS modulation is given by
\vspace{-2.5mm}
\begin{equation}\label{IOOTFS}
Y_{\mathrm{DD}}\hspace{-0.5mm}\left[ {l,k} \right] \hspace{-1mm}=  \hspace{-1mm}\sum\limits_{k' = 0}^{N - 1} \hspace{-0.5mm}{\sum\limits_{l' = 0}^{M - 1} \hspace{-1mm}{X_{\mathrm{DD}}\hspace{-0.5mm}\left[ {l',k'} \right]} } {H}_{\mathrm{DD}}\hspace{-0.5mm}\left[ {l,k, l',k'} \right] + Z_{\mathrm{DD}}\left[l,k\right],\vspace{-3mm}
\end{equation}
where $Z_{\mathrm{DD}}\left[l,k\right]$ denotes the effective DD domain noise.
The discrete DD domain channel ${H}_{\mathrm{DD}}\left[ {l,k, l',k'} \right]$ is given by
\vspace{-3mm}
\begin{equation}\label{EffectiveDDDomainChannel}
{H}_{\mathrm{DD}}\left[ {l,k, l',k'} \right] \hspace{-1mm}= \hspace{-1mm}\sum\nolimits_{i=1}^{P} {h_i} w\left({l,k, l',k',l_{i},k_{i}}\right) {e^{ - j2\pi {\nu_i}{\tau_i}  }},\vspace{-2mm}
\end{equation}
where $k_{i} = {\nu_i}NT \in \mathbb{R}$ and $l_{i} = {\tau_i} M\Delta f\in \mathbb{R}$.
We distinguish between \textit{integer} and \textit{fractional} Doppler and delay scenarios based on whether $k_{i}$ and $l_{i}$ are integer or fractional numbers, respectively\cite{RavitejaOTFS}.
In \eqref{EffectiveDDDomainChannel}, $w\left({l,k, l',k',l_{i},k_{i}}\right)$ is the sampling function for analog DD domain channel response, which captures joint effects of delay and Doppler shifts, TX pulse shaping filter, and RX filter, and is given by \eqref{WindowDDDomainRectPulse} at the top of next page, where the ambiguity function ${{A_{{g_{\rm tx}}{g_{ \rm rx}}}}\left(\tau,\nu \right)}$ is defined as
\setcounter{equation}{9}
\vspace{-2mm}
\begin{equation}\label{AmbiguityFunction}
{{{A_{{g_{\rm tx}}{g_{\rm rx}}}}\left(\tau,\nu \right)}} = \int_{-\infty}^{\infty} {g_{{\rm{tx}}}}\left( {t} \right)g_{{\rm{rx}}}^ * \left( {t - \tau } \right){e^{ - j2\pi \nu \left( {t - \tau } \right)}}\mathrm{d}t.\vspace{-3mm}
\end{equation}

For the widely adopted rectangular TX and RX filters and assuming integer $k_{i}$ and $l_i$\cite{RavitejaOTFS}, \eqref{IOOTFS} simplifies to the well-known 2D circular convolution input-output relationship as follows:
\vspace{-2mm}
\begin{align}\label{DDIOModel_Rect_Integer}
Y_{\mathrm{DD}}\left[ {l,k} \right]& =	\sum\nolimits_{i=1}^{P}h_i  X_{\mathrm{DD}}\left[\left(l\hspace{-0.5mm}-\hspace{-0.5mm}l_{i}\right)_M,\left(k\hspace{-0.5mm}-\hspace{-0.5mm} k_{i}\right)_N\right]\alpha\left[l,k,l_i,k_i\right]\notag\\[-1mm]
& + Z_{\mathrm{DD}}\left[l,k\right],
\end{align}
\setcounter{equation}{12}
\vspace{-8mm}
\par
\noindent
where $\alpha\left[l,k,l_i,k_i\right]$ is a phase rotation term defined in \eqref{PhaseRotation} at the top of next page.

\vspace{-4mm}
\subsection{DZT-based OTFS}

\begin{figure}[t]
	\centering
	\includegraphics[width=3.2in]{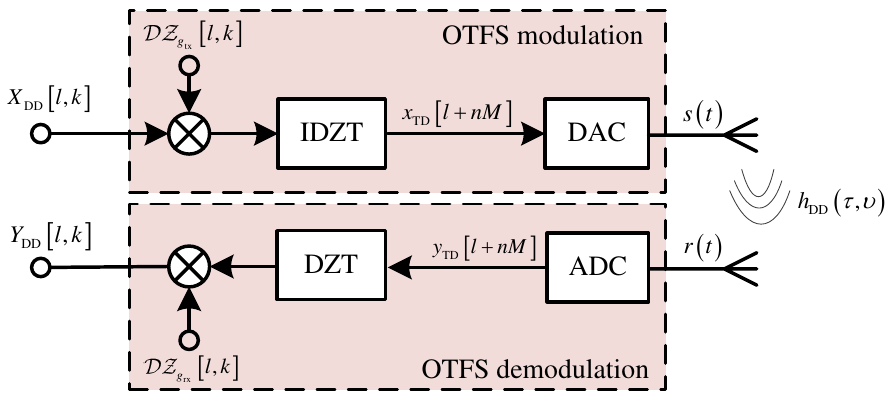}\vspace{-3mm}
	\caption{DZT-based OTFS transceiver architecture.}\vspace{-6mm}
	\label{DZT_Based_OTFS}%
\end{figure}

SFFT-based OTFS modulation maps the DD domain information symbols first to the TF domain and then to the time domain\cite{RavitejaOTFS,FarhangCPOFDM}.
In contrast, DZT-based OTFS modulation directly transforms the DD domain information symbols into the time domain, as shown in Fig. \ref{DZT_Based_OTFS}\cite{MohammedZTOTFS,lampel2021orthogonal}.
The DZT and inverse DZT (IDZT) define a mapping between a one-dimensional (1D) sequence and a 2D sequence. Their definitions are provided in the following.

\begin{Def}[Discrete Zak Transform]
	Let $x$ be a periodic sequence with period $MN$. The DZT of $x$ is defined as
	\vspace{-2mm}
	\begin{equation}\label{DZT}
	\mathcal{DZ}_x\left[l,k\right] \buildrel \Delta \over = \frac{1}{\sqrt{N}}\sum\nolimits_{n=0}^{N-1}x\left[l+nM\right]e^{-j2\pi\frac{n}{N}k},\vspace{-2mm}
	\end{equation}
	for $l \in \{0,\ldots,M-1\}$ and $k \in \{0,\ldots,N-1\}$.
\end{Def}

\begin{Def}[Inverse Discrete Zak Transform]
	Given the $MN$-periodic sequence $x$, whose DZT is given by $\mathcal{DZ}_x\left[l,k\right]$, the IDZT is defined as
	\vspace{-2mm}
	\begin{equation}\label{IDZT}
	x\left[l+nM\right] \buildrel \Delta \over = \frac{1}{\sqrt{N}}\sum\nolimits_{k=0}^{N-1}\mathcal{DZ}_x\left[l,k\right]e^{j2\pi\frac{n}{N}k},\vspace{-2mm}
	\end{equation}
	for $l \in \{0,\ldots,M-1\}$ and $n \in \{0,\ldots,N-1\}$.
\end{Def}

As shown in \cite{lampel2021orthogonal}, a time-delay (TD) domain sequence can be obtained by applying the IDZT to the DD domain symbols at the OTFS TX, i.e., 
\vspace{-2mm}
\begin{equation} \label{IDZT}
{x}_{\mathrm{TD}}\left[ {l+nM} \right] \hspace{-0.5mm}=\hspace{-0.5mm} \sqrt{M}\sum\nolimits_{k = 0}^{K - 1} {X_{\mathrm{DD}}\left[ {l,k} \right] \mathcal{DZ}_{g_{\mathrm{tx}}}\left[l,k\right] {e^{j2\pi \frac{n}{N}k}}},\vspace{-1mm}
\end{equation}
where $\mathcal{DZ}_{g_{\mathrm{tx}}}\left[l,k\right]$ is the DZT of the \textit{periodically extended} sampled TX pulse shape $g_{\mathrm{tx}} \left(\frac{[k]_{MN}}{M}T\right)$, $k \in \mathbb{Z}$.
The OTFS transmit signal $s(t)$ can be obtained by DAC of ${x}_{\mathrm{TD}}\left[ {l+nM} \right]$.
Interpolating with a rectangular kernel, the OTFS transmit signal is given by
\vspace{-2mm}
\begin{equation}\label{DAC}
s(t) \hspace{-0.5mm}= \hspace{-0.5mm} {x}_{\mathrm{TD}}\left[ {l\hspace{-0.5mm}+\hspace{-0.5mm}nM} \right], t\in \left[\frac{l\hspace{-0.5mm}-\hspace{-0.5mm}0.5}{M}T\hspace{-0.5mm}+\hspace{-0.5mm}nT,\frac{l\hspace{-0.5mm}+\hspace{-0.5mm}0.5}{M}T\hspace{-0.5mm}+\hspace{-0.5mm}nT\right].\vspace{-1.5mm}
\end{equation}

At the OTFS RX, the channel output $r(t)$ in \eqref{LTVChannel} is subjected to an analog-to-digital conversion (ADC) and the resulting TD domain received signal is given by
\vspace{-2mm}
\begin{equation}\label{ADC}
y_{\mathrm{TD}}\left[ {l+nM} \right] = r(t)|_{t = \frac{l}{M}T+nT}.\vspace{-2mm}
\end{equation}
Then, the DD domain received signal is obtained by performing DZT on the TD domain received signal, i.e., 
\vspace{-2mm}
\begin{equation}\label{DZT}
Y_{\mathrm{DD}}\left[ {l,k} \right] = \sqrt{MN}\mathcal{DZ}_{y_{\mathrm{TD}}}\left[l,k\right] \mathcal{DZ}^*_{g_{\mathrm{rx}}}\left[l,k\right],\vspace{-1.5mm}
\end{equation}
where $\mathcal{DZ}_{y_{\mathrm{TD}}}\left[l,k\right]$ is the DZT of $y_{\mathrm{TD}}\left[ {l+nM} \right]$ and $\mathcal{DZ}_{g_{\mathrm{rx}}}\left[l,k\right]$ is the DZT of the \textit{periodically extended} sampled RX filter $g_{\mathrm{rx}} \left(\frac{[k]_{MN}}{M}T\right)$, $k \in \mathbb{Z}$.

A common choice for the TX and RX filters are rectangular filters.
In this case, we have $\mathcal{DZ}_{g_{\mathrm{tx}}}\left[l,k\right] = \mathcal{DZ}_{g_{\mathrm{rx}}}\left[l,k\right] = \frac{1}{\sqrt{MN}}$, $\forall l,k$.
Moreover, assuming both integer delays and integer Doppler shifts, the DD domain input-output relationship can be obtained as \cite{lampel2021orthogonal}
\vspace{-3mm}
\begin{align}\label{IOZakDomain}
Y_{\mathrm{DD}}\left[ {l,k} \right] &=\sum\nolimits_{i=1}^{P}h_i  X_{\mathrm{DD}}\left[\left(l\hspace{-0.5mm}-\hspace{-0.5mm}l_{i}\right)_M,\left(k\hspace{-0.5mm}-\hspace{-0.5mm} k_{i}\right)_N\right]\alpha\left[k,l,k_i,l_i\right]\notag\\[-1mm]
& + Z_{\mathrm{DD}}\left[l,k\right],
\end{align}
\vspace{-6mm}
\par
\noindent
which is identical to \eqref{DDIOModel_Rect_Integer}.
We refer to \cite{lampel2021orthogonal} for a more detailed discussion of the input-output relationship for non-rectangular TX and RX filters, fractional delays, and fractional Doppler shifts. 

\vspace{-4mm}
\subsection{Their Similarities and Differences}
{The SFFT- and DZT-based OTFS implementations are both suitable for high-mobility communications.
Besides, for rectangular TX and RX filters, integer delays, and integer Doppler shifts, SFFT- and DZT-based OTFS lead to the same input-output relationship, see \eqref{DDIOModel_Rect_Integer} and \eqref{IOZakDomain}.
Unveiling their connections for the case of non-rectangular filters, fractional delays, and fractional Doppler shifts is still an open problem in the literature. 

SFFT-based OTFS is compatible with conventional OFDM by concatenating the ISFFT/SFFT module and an OFDM modulator/demodulator, while DZT-based OTFS enjoys a lower computational complexity as only $N$-point inverse finite Fourier transform (IFFT) and finite Fourier transform (FFT) are needed with a symbol-wise interleaver/de-interleaver at the OTFS transceivers, respectively, see \eqref{IDZT} and \eqref{DZT}.
Moreover, the SFFT-based OTFS architecture facilitates TF domain signal processing, such as window function design\cite{WeiWindowOTFS} and design of multiple access schemes\cite{KhammammettiMA}.
On the other hand, for DZT-based OTFS, the TX and RX filters can be implemented digitally in the DD domain based on their DZTs.
This not only reduces the implementation complexity, but also facilitates a systematic pulse shape design in the DD domain.
A detailed discussion on
pulse shape design can be found in Part II of this tutorial.}

\vspace{-2mm}
\section{Discussions on OTFS Modulation}
Based on the transceiver architectures introduced in Section II, we shed light on some fundamental concepts of OTFS modulation in this section.
\vspace{-4mm}
\subsection{What is the Delay-Doppler Domain}
In the context of signal processing, we define and describe a signal in a certain domain.
Describing a signal in the time domain characterizes the fluctuation of its amplitude and phase w.r.t. the time variable, while representing a signal in the frequency domain characterizes the corresponding fluctuation rate.
Furthermore, the TF domain is expanded by the time duration and frequency bandwidth occupied by a signal.
The characterization of a signal in the TF domain reveals the temporal evolution of its spectral content (within a certain time duration).
On the other hand, the delay and Doppler parameters are usually used to characterize a time-varying system rather than a signal.
For example, $h\left(\tau,\nu\right)$ in \eqref{DDDomainChannel} characterizes the response of a time-varying system that introduces delays $\tau_i$, $\forall i$, and Doppler shifts $\nu_i$, $\forall i$, to the input signal.
OTFS modulation is based on signal representation theory\cite{Hadani2018otfs}, cf. \eqref{ISFFT}, \eqref{MulticarrierMod}, and \eqref{IDZT}, which allows to characterize a signal directly in the DD domain.
In particular, a DD domain impulse $\delta\left[l-l_0,k-k_0\right]$, $l,l_0 \in \{0,\ldots,M-1\}$, $k,k_0 \in \{0,\ldots,N-1\}$ corresponds to a \textit{time-varying pulsetone} in the time domain\cite{Hadani2018otfs}, since it can be viewed as the output of a \textit{virtual time-varying system} that introduces delay $l_0\frac{1}{M\Delta f}$ and Doppler shift $k_0\frac{1}{NT}$ w.r.t. an input DD domain impulse $\delta\left[l,k\right]$.
At the OTFS RX, when transforming a signal from the time or TF domain to the DD domain, cf. \eqref{MulticarrierDeMod}, \eqref{SFFT}, and \eqref{DZT}, we are inverting the channel operations the TX signal has experienced.
{According to the uncertainty principle of the Fourier transform, the longer the TX signal duration, the higher the resolution for distinguishing different Doppler shifts in the channel, while the larger the TX signal frequency bandwidth, the higher the resolution for distinguishing different delay shifts in the channel \cite{MohammedZTOTFS}.}

\vspace{-4mm}
\subsection{How to Embed Information into the DD Domain}
%
As shown in \eqref{IDZT}, the OTFS TX modulates information symbols $X_{\mathrm{DD}}\left[ {l,k} \right]$ onto the DD domain pulse samples $\mathcal{DZ}_{g_{\mathrm{tx}}}\left[l,k\right]$, $l \in \{0,\ldots,M-1\}$, $k \in \{0,\ldots,N-1\}$, which can be interpreted as the generation of a set of \textit{{time-varying carrier signals}} in the time domain to carry the corresponding information symbol.
A set of 2D impulses with zero bandwidth in the DD domain, i.e., $\delta\left[l-l_0,k-k_0\right]$, $l,l_0 \in \{0,\ldots,M-1\}$, $k,k_0 \in \{0,\ldots,N-1\}$, would be ideal for carrying information since they are perfectly localized\cite{MohammedZTOTFS}.
Unfortunately, this is not possible in practice since such impulses would require infinite time duration and infinite frequency bandwidth.
As shown in \cite{MohammedZTOTFS}, 2D sinc pulses (their samples are given by $\mathcal{DZ}_{g_{\mathrm{tx}}}\left[l,k\right]$ in \eqref{IDZT}, $l \in \{0,\ldots,M-1\}$, $k \in \{0,\ldots,N-1\}$) with finite time duration and frequency bandwidth, which are approximately orthogonal in the DD domain, constitute an efficient set of DD domain information-carrying impulses.

\vspace{-4mm}
\subsection{Benefits of DD Domain Modulation}
{Due to the Heisenberg uncertainty principle, it is impossible to find a 2D signal perfectly localized in the TF domain, while in the DD domain, we can obtain a set of 2D localized signals (their samples are given by $\mathcal{DZ}_{g_{\mathrm{tx}}}\left[l,k\right]$ in \eqref{IDZT}) that are approximately orthogonal \cite{MohammedZTOTFS}.
Moreover, as mentioned before, the OTFS modulator employs time-varying carrier signals for carrying information, which caters to the time-varying features of high-mobility channels.
In particular, the high-mobility channel, corresponding to DD domain channel response, $h_{\mathrm{DD}}\left(\tau,\nu\right)$ in \eqref{DDDomainChannel} can be modeled as a linear time-varying system concatenated with the \textit{virtual time-varying system} formed by the OTFS modulator.
As a result, the high-mobility channel \textit{additively} changes the time-varying features of the carrier signals, i.e., the received carrier signal experiences a shorter or a longer delay ($l\pm l_i$)  and a lower or a higher Doppler shift ($k\pm k_i$), compared to the transmitted carrier signal in the $\left(l,k\right)$-th DDRE.}
As a result, the additive channel operations on the carrier signals, i.e., the delay and Doppler shifts, can be simply described by a shift operation and a phase rotation in the DD domain, i.e., \eqref{DDIOModel_Rect_Integer} and \eqref{IOZakDomain}, which is beneficial for efficient and reliable channel equalization.

\vspace{-2mm}
\section{Challenges Ahead}
So far, the research on OTFS modulation has focused on the characterization of the input-output relationship, transceiver design, and potential synergies with other communication and signal processing techniques, as elaborated in Parts I, II, and III, respectively. 
However, some aspects of OTFS modulation are not well understood or have not been addressed by the research community, yet.
In this section, we discuss important open problems, which might be the main challenges ahead for OTFS research.

\vspace{-4mm}
\subsection{Latency Concern}
Compared to OFDM, OTFS modulation spreads information over a longer block to combat channel fluctuations, which may increase the overall latency. 
In particular, the duration of each OTFS signal is $N$ times of that of an OFDM symbol, with the same number of subcarriers and the same subcarrier spacing\cite{RavitejaOTFS}. 
Fortunately, multiplexing information in the DD domain turns a time-varying channel into a quasi-static channel in the DD domain, which might significantly reduce the amount of signaling overhead required for channel estimation.
A comprehensive comparative study of the latency of OTFS and OFDM is still missing.
Besides, the latency of OTFS can be made identical to that of OFDM by increasing the subcarrier spacing or reducing $N$ at the expense of a reduced Doppler resolution.
A detailed performance comparison of OTFS and OFDM for a given latency is an interesting research direction.

\vspace{-4mm}
\subsection{Fractional Doppler}
To limit the latency of OTFS modulation, the signal duration $NT$ cannot be chosen exceedingly large, which gives rise to the well-known fractional Doppler issue\cite{RavitejaOTFS}. 
Existing works in the literature have addressed the DD domain channel estimation problem in the case of fractional Doppler by various means, e.g., using windowing\cite{WeiWindowOTFS}, oversampling\cite{YaoOversampling}, and off-grid estimation\cite{ZhiqiangOffGrid}.
However, detection performance may significantly degrade in the case of fractional Doppler. 
In fact, the commonly used message passing-based detection algorithms \cite{RavitejaOTFS} are prone to diverge, due to the loops in the factor graph caused by fractional Doppler shift.
How to improve detection performance and robustness in the case of fractional Doppler is a critical issue.
{We note that the authors in \cite{ShuangyangCrossDomain2021,YuanOTFS} have proposed a cross domain message passing detector and a variational Bayes detector, respectively, which can improve the convergence performance in the case of fractional Doppler.}

\vspace{-4mm}
\subsection{Multiple Access}
Serving multiple users in high-mobility environments is a challenging task.
For OTFS modulation, it is not clear yet in which domain the user should be multiplexed.
Conventional TF domain multiple access schemes may suffer from severe multi-user interference (MUI) caused by doubly selective fading channels.
On the other hand, DD domain multiple access requires careful allocation of the DDREs and the guard space between users to minimize MUI.
Moreover, the 2D circular convolution input-output relationship of OTFS leads to special MUI patterns that might be exploited for user scheduling and resource allocation design.
Furthermore, compared to orthogonal multiple access, power domain and/or code domain non-orthogonal multiple access schemes \cite{wei2018multibeam} can accommodate more users while imposing a critical challenge for interference management.

\vspace{-4mm}
\subsection{MU MIMO-OTFS}
Multiple-input multiple-output (MIMO) systems can provide additional spatial DoFs for OTFS systems, which may be exploited to improve communication reliability and the achievable rate. 
Moreover, these additional DoFs can be used to accommodate multiple users in the spatial domain, which would allow to avoid the guard resource elements needed in existing DD and TF domain multiple access schemes\cite{RavitejaOTFSCE}.
However, incorporating OTFS modulation in MU MIMO systems is a challenging task since the signals have to be optimized simultaneously in the delay, Doppler, and spatial domains.
The authors of \cite{PandeyMIMOOTFS} investigated the precoding design for massive fully-digital MIMO-OTFS systems, which may incur exceedingly high hardware complexity and power consumption.
In fact, a general and concise MU MIMO-OTFS model, based on which channel estimation,  precoding, and multiple access schemes can be designed, has not been reported so far. 

\vspace{-4mm}
\subsection{Cross-layer Design}
High-mobility communication networks, such as UAV and V2V networks, may suffer not only from time-varying channels but also from a highly dynamic network topology.
However, existing works on OTFS modulation mainly focus on the physical layer design to improve communication performance. 
Cross-layer design spanning both the physical and network layers for OTFS systems is a critical but challenging task.
In fact, cross-layer design facilitates coordination, interaction, and joint optimization of communication protocols across different layers, which is beneficial for mobility management and energy-efficient and reliable communications.
In high-mobility environments, it is critical to sense the fast time-varying network state information (NSI), such as the location and velocity of the network elements exploiting the OTFS modulation waveform, and to design the OTFS transceivers accordingly.
Furthermore, the network topology may be actively adjusted depending on the state of the high-mobility environment to improve the communication performance.

\vspace{-2mm}
\section{Conclusions}
As the first part of this three-part tutorial, this letter presented and compared two common implementations of OTFS, namely SFFT- and DZT-based OTFS.
Based on the introduced OTFS transceiver architectures, some fundamental concepts of OTFS modulation were discussed.
Subsequently, several important research challenges regarding OTFS modulation were highlighted. 
Part II and Part III will discuss OTFS transceiver design and the benefits of OTFS for ISAC and other applications, respectively.
\bibliographystyle{IEEEtran}
\bibliography{OTFS}

\end{document}